\documentclass[twoside,10pt]{article}
\usepackage{amsmath,amssymb}
\usepackage{tam13}
\usepackage{xspace}
\usepackage{caption}
\usepackage{setspace}
\usepackage{color}

\title{Micro Black Hole Production and Evaporation}
\shorttitle{Micro Black Hole}
\authors{S.~Nafooshe\email{Saeede.Nafooshe@ung.si}, M.J.O'Loughlin, M.V.Garzelli}
\shortauthors{S.~Nafooshe and et al.}
\affiliations{
 University of Nova Gorica, Vipavska 13, SI-5000 Nova~Gorica, Slovenia
}

\abstract{
It has been conjectured that Micro Black Holes (MBH) may be formed in the presence of large extra
dimensions. These MBHs have very small mass and they decay almost instantaneously. Taking into
consideration quantum effects, they should Hawking radiate mainly to Standard Model particles, this
radiation then gets modified by the non trivial geometry around the MBHs; the so called greybody factors
which filter the Hawking radiation. To test the validity of MBH models, one needs to investigate it
experimentally. A primary tool in this investigation is simulation of the MBH formation and evaporation,
including all theoretical work that has been performed up to now. BlackMax and CHARYBDIS2 are the most
modern and realistic simulators currently available. However they still suffer from a lack of important
parameters. In this article we will discuss the primary work that we have done to study the possible changes that
can be implemented in the simulations.
}

\begin{document}
\maketitle

\section{Introduction}
 In the Standard Model (SM) of particle physics the electroweak scale M$_\text{EW}$ is set by the Higgs potential and is around $1$TeV. If the SM is valid up to Planck mass  M$_\text{Pl}$$\thicksim 10^{16}$\,TeV, one expects fine tuning between these two scales. However there is a big difference between them. This big gap between these two fundamental scales is known as the hierarchy problem. Several scenarios (supersymmetry solution, existence of extra dimensions, etc.) in different contexts have been proposed to provide a potential resolution of this problem.\\ One of these models is the "large extra dimensions" model that was introduced by Arkani, Dimopoulos and Dvali (ADD model) in $1998$ \cite{r:ADD}. In this model all the SM particles are confined to a 3-dimensional brane that is located in a higher dimensional bulk. The graviton can propagate in the compactified $n$-extra spatial dimensions through the bulk. The 4-dimensional Planck scale and the $4+n$-dimensional gravity scale $M_{D}$ are related via
\begin{equation}
M_\text{Pl}^2=V_{n}M_{D}^{2+n},
\label{e:add}
\end{equation}
where $V_{n}$ is the volume of the compactified dimensions. The ADD model assumes that $M_{D}$ is of the order of a few TeV and with this assumption the hierarchy between $M_{D}$ (the fundamental gravity scale) and M$_\text{EW}$ thanks to the large volume of the higher dimensional space can be resolved.
\\
 If nature realizes this model MBH formation at TeV scale is one of the consequence of such a model. MBHs might be created at the Large Hadron Collider (LHC) \cite{r:lhc} or by the collision of ultra high energy cosmic rays (UHCRs) with the Earth's atmosphere \cite{r:cosmic}. In contrast with normal black holes that have a long life time, MBHs have a very short life time of order of $10^{-26}$\,s. This means that once they will be created they will evaporate instantaneously. All process (for MBHs with mass\,$\gtrsim \,10M_{D}$) of the MBH formation and evaporation can be classified into four stages: balding phase, spindown phase, Schwarzschild phase and Planck phase. During the first stage the MBH loses its hair mainly by gravitational emission and in the second and third phase Hawking radiation \cite{r:haw} takes mass and angular momentum of the MBH away.\\ The Hawking emission of particles has a thermal spectrum and could be a detectable sign of the evaporation of MBHs. Due to the effect of curved space time around MBH the spectrum a long way from the MBH is different from the black body radiation due to the so called grey body factors. These signals have been simulated by the MBH event generators like BlackMax \cite{r:BlackMax} and CHARYBDIS2 \cite{r:charybdis}. These generators include all the grey body factors which are known up to now. However grey body factors for graviton emission from a rotating black hole have not yet been calculated and included in these codes.\\
 In this article we study the sensitivity of BlackMax to the exact form of these grey body factors with the purpose of understanding the required accuracy for an approximation to gravity grey body factors.

\section{MBH formation}

In hadron-hadron collisions, MBH formation can take place when two partons with associated spin and mass come sufficiently close to each other such that they form a small region with large amount of energy that can be surrounded by a hoop Schwarzschild radius, R$_\text{s}$, of the energy \cite{r:hoop}.\\
The mass of this formed MBH is approximately equal to center of mass energy of the collision and its angular momentum can be explained in terms of the impact parameter and the mass of the MBH. The total cross section from geometrical arguments is approximately equal to the interaction area, $\pi R_{\text{s}}^{2}$.\\

\subsection{Decay process}
\label{ss:dp}
The MBH decay takes place in four stages \cite{r:stage}. The first stage is balding phase. Black hole has no hair; it quickly becomes bald by shedding out the multiple moments by emission of gravitational and electromagnetical radiation. This can increase its spin. The next stage is the spin down phase. During this phase the MBH emits Hawking radiation and preferentially it sheds its angular momentum until it settles down into a stationary Schwarzschild MBH. In the Schwarzschild phase, it continues to Hawking radiate and loses its mass. One can compute the power spectrum and relative emission rates for this process. It loses its mass until it reaches the Planck scale where $M_{\text{MBH}}\sim M_{D}$. To describe this phase a full quantum gravitational description is required and due to the lack of a theory of quantum gravity the Planck phase is the most unknown phase in the decay process (we will not discuss it any more in the following).

\section{Hawking radiation}
 Black holes are thermal objects, they have a specific temperature and entropy and they can radiate. This radiation is known as Hawking radiation. Hawking showed that exactly at the event horizon, the emission rate of a black hole in a mode with frequency $\omega$ is given by
\begin{equation}
\Gamma(\omega)\text{d}\omega=\frac{1}{e^{\beta \omega}\pm 1}\frac{\text{d}^{3}k}{(2\pi)^{3}},
\label{e:blackbody}
\end{equation}
where $\beta$ is the inverse of Hawking temperature and the plus and minus sign stands for fermion/boson respectively. It is clear that at the horizon the spectrum of radiation is a black body spectrum and it is perfectly thermal. However the observer that is located very far from the black hole will not measure a black body spectrum but a grey body spectrum.

\subsection{Grey body factors}
\label{ss:grey}
The geometry of the space-time surrounding a black hole is curved. This geometry is encoded in the potential $V(r)$ and this potential acts as a filter; a part of the radiation is transmitted and travels to infinity, whereas the other part is reflected back into the black hole. The spectrum that can be observed at infinity is
\begin{equation}
\Gamma(\omega)\text{d}\omega=\frac{\gamma(\omega)}{e^{\beta \omega}\pm 1}\frac{\text{d}^{3}k}{(2\pi)^{3}},
\label{e:greybody}
\end{equation}
where $\gamma(\omega)$ is the grey body factor. It is energy dependent and also depends on the frequency of the particles under consideration. This factor can be defined in terms of the transmission and reflection coefficients of the potential barrier.\\
In the case of higher dimensional rotating MBHs whose gravitational background can be described by the Myers-Perry solution \cite{r:MP}, the energy emission rate is,
\begin{equation}
\frac{\text{d}E}{\text{d}t}=\frac{1}{2\pi}\sum_{\ell,m}|A_{\ell,m}|^{2} \frac{\omega\, \text{d}\omega}{\exp((\omega-m\Omega)/T_{\text{H}})\mp 1},
\label{e:MPgrey}
\end{equation}
where $|A_{\ell,m}|^{2}$ is the grey body factor and $T_{\text{H}}$ is the Hawking temperature
\begin{equation}
T_{\text{H}}=\frac{(n+1)+(n+1)a_{\ast}^{2}}{4\pi(1+a_{\ast}^{2})r_{\text{h}}},
\label{e:th}
\end{equation}
and $\Omega$ is the angular velocity which is defined by
\begin{equation}
\Omega=\frac{a_{\ast}}{(1+a_{\ast}^{2})r_{\text{h}}},
\label{e:omega}
\end{equation}
where $a_{\ast}$ is the spin parameter and r$_{\text{h}}$ is horizon radius.\\
\begin{figure}[h!]
\centering
\parbox{5.25cm}{
\hspace{-1cm}\includegraphics[width=6cm]{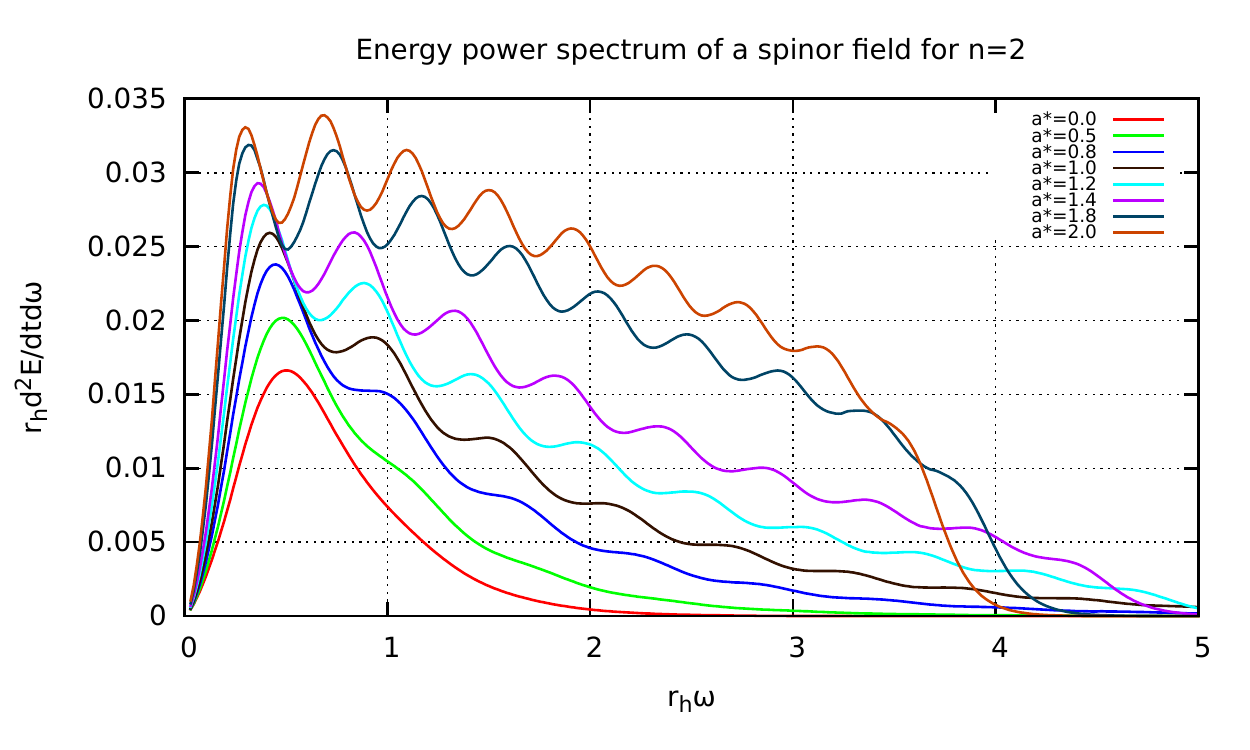}
\label{fig:2figsA}}
\qquad
\hspace{-0.95cm}\begin{minipage}{5cm}
\includegraphics[width=6cm]{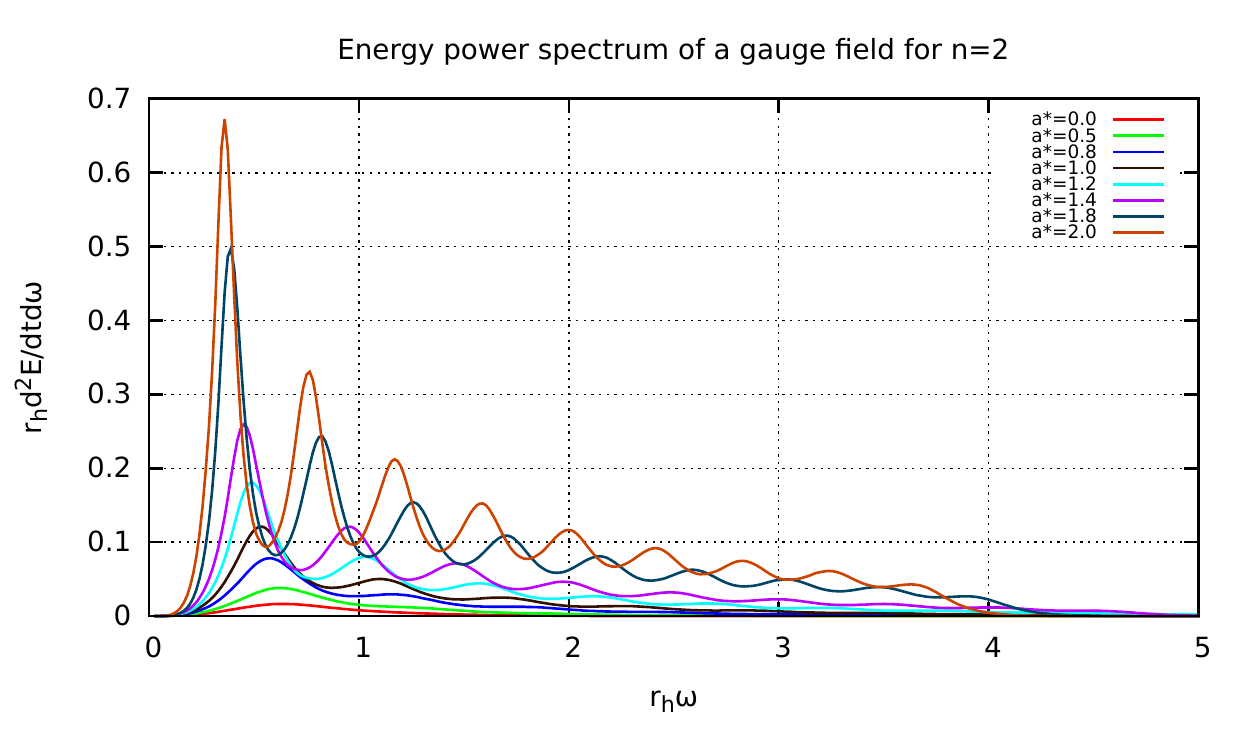}
\label{fig:2figsB}
\includegraphics[width=6cm]{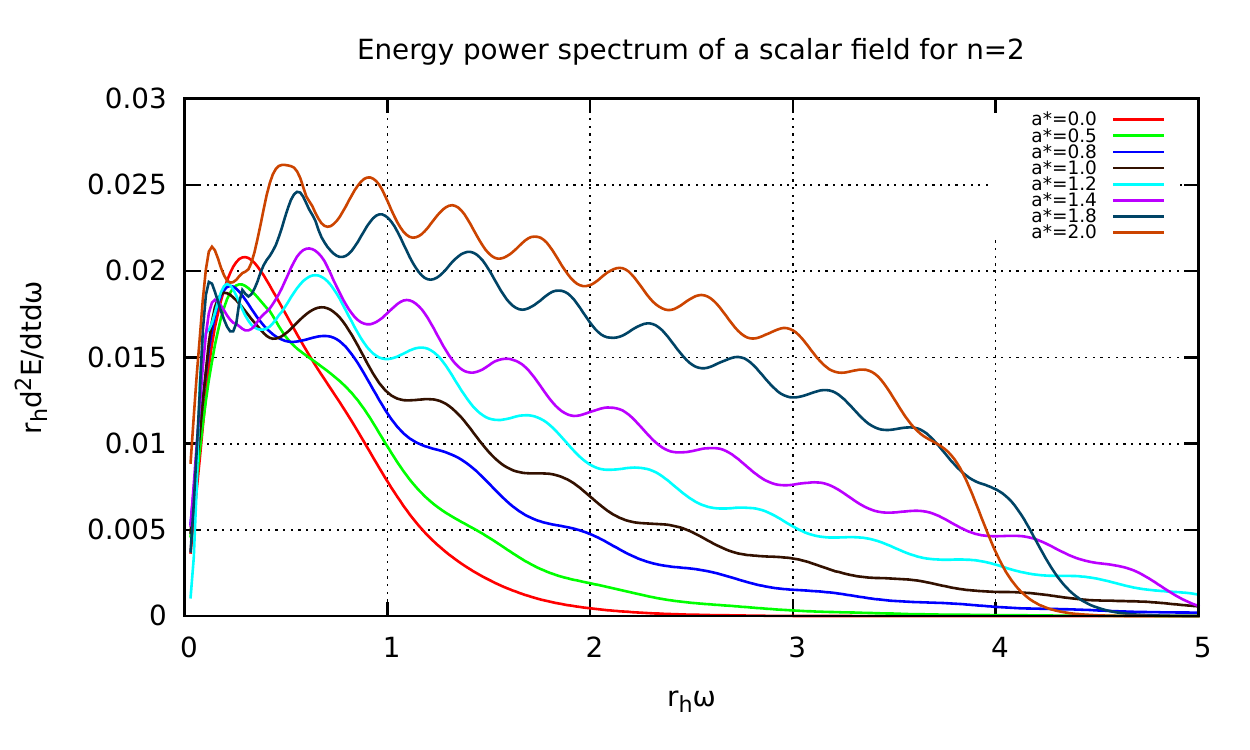}
\label{fig:2figsc}
\end{minipage}
\caption{Energy power spectrum for spinor, gauge and scalar fields for $n=2$.}
\label{f:ep}
\end{figure}
The energy power spectrum has been presented for different spin parameters in the figure \ref{f:ep} for spinor, gauge and scalar fields in the presence of two extra dimensions.

\section{Micro black hole event generators}
In recent years several MBH event generators have been developed for simulating the formation and decay of MBHs formed in high energy collisions. Among them, CHARYBDIS2 and BlackMax are the most recent ones, and they include all the grey body factors which are known up to now as well as the effect of rotation has been taken into account in these two generators.\\
For simulation of the following parton shower and hadronization process they can be interfaced to PYTHIA \cite{r:pythia} or HERWIG \cite{r:herwig} Monte Carlo event generators.

\subsection{BlackMax sensitivity to the change of greybody factors}
\label{subsec:BS}
\begin{figure}[h!]
\centering
\parbox{5.25cm}{
\hspace{-0.85cm}\includegraphics[width=6cm]{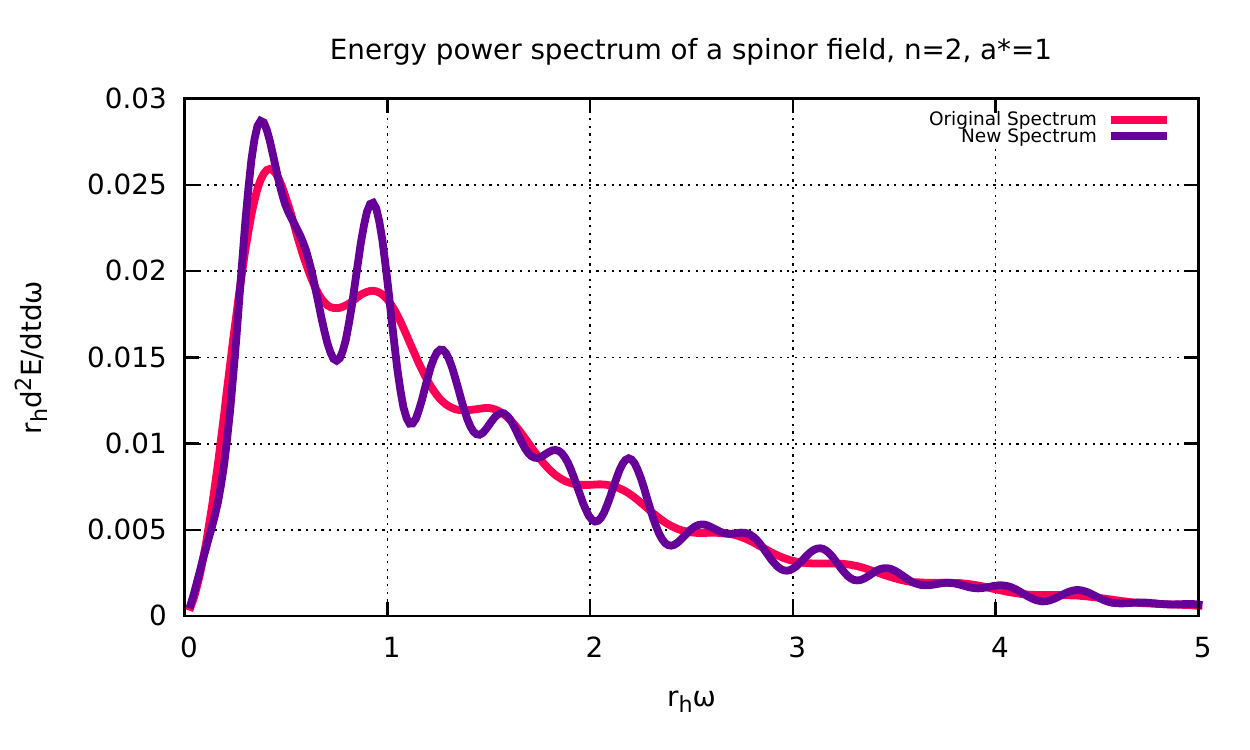}
\caption*{(a)}
\label{fig:2figsA}}
\qquad
\hspace{-0.85cm}\begin{minipage}{5cm}
\includegraphics[width=6cm]{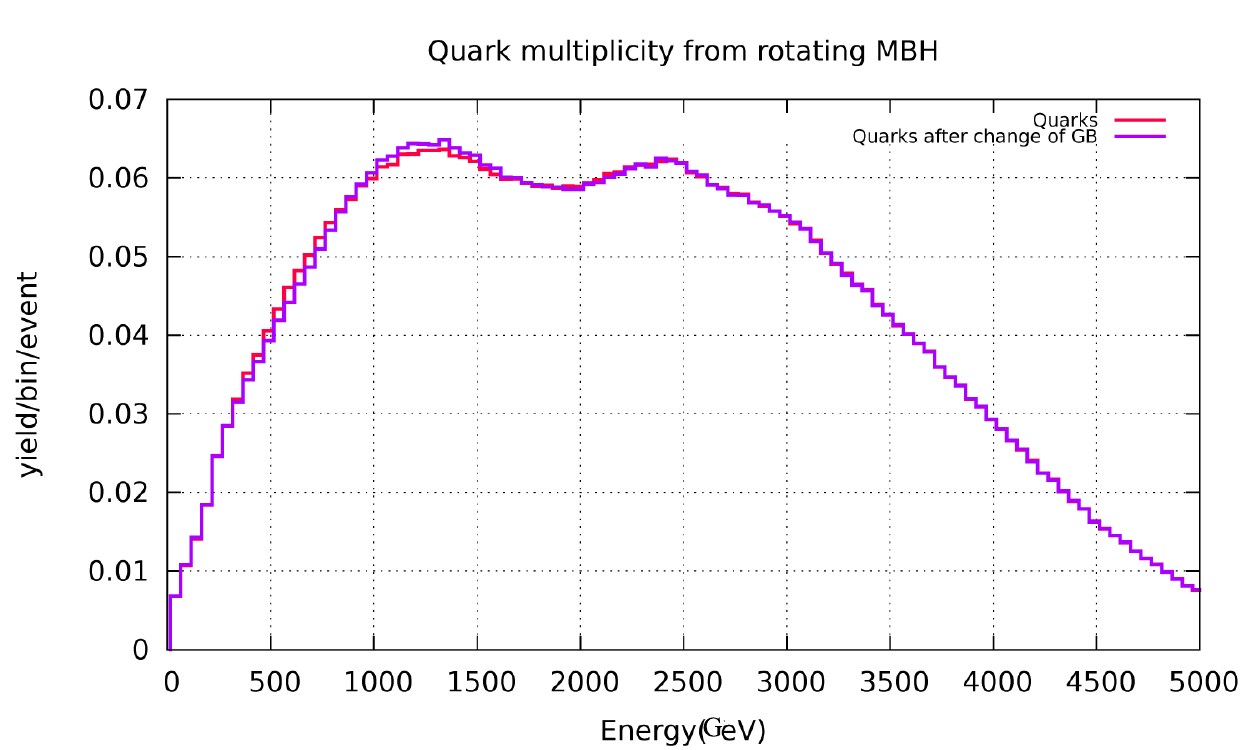}
\caption*{(b)}
\label{fig:2figsB}
\includegraphics[width=6cm]{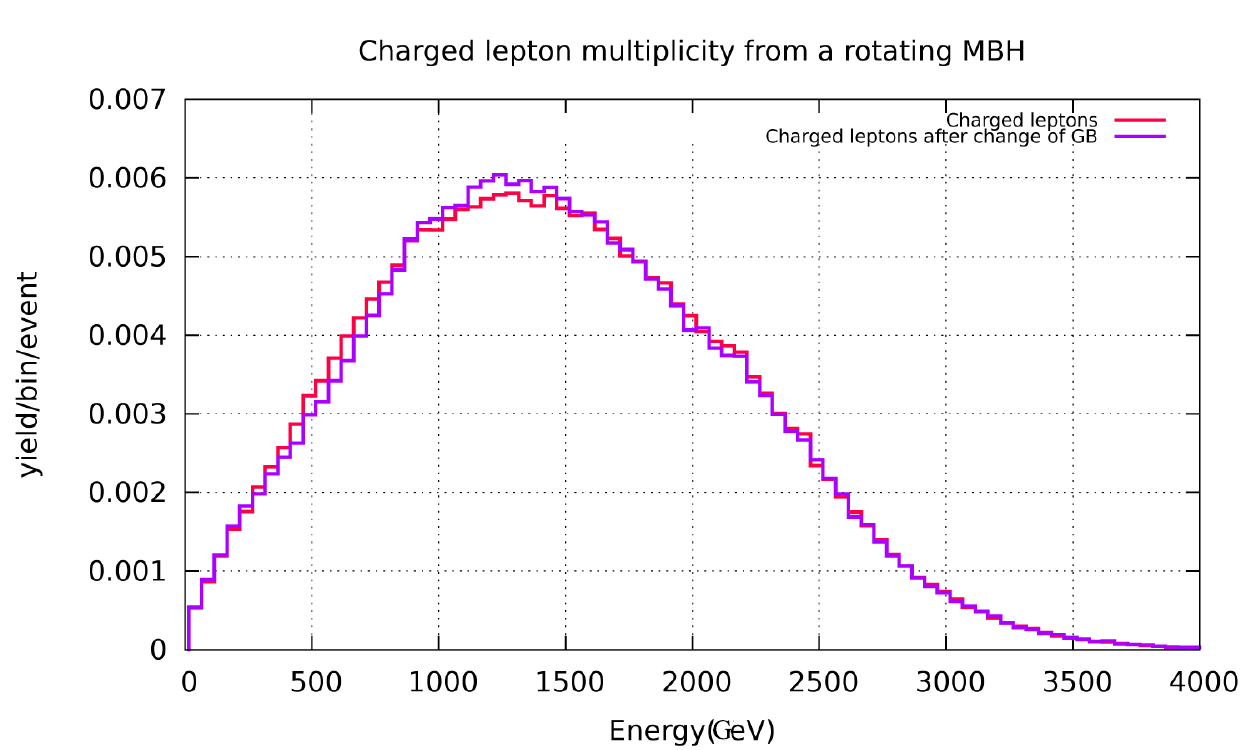}
\caption*{(c)}
\label{fig:2figsc}
\end{minipage}
\caption{(a) Energy power spectrum from a spinor field for $a_{\ast}=1$ before and after change of greybody factors., (b) and (c) quark and charged lepton multiplicities before and after change of greybody factors.  }
\label{f:group}
\end{figure}

As we mentioned before, the probability for the emission of each degree of freedom is given by the greybody factor for the specific mode. These factors depend on the properties of the emitted particles (charge, spin, mass, momentum) and the MBH (mass, spin, charge) and also on environmental properties like the location of the MBH with respect to the brane and the number of extra dimensions \cite{r:BlackMax}. It is important to know about the sensitivity of the MBH event generators to the exact form of the greybody factors.\\ 
\begin{figure}[h!]
\centering
\parbox{5.25cm}{
\hspace{-0.85cm}\includegraphics[width=6cm]{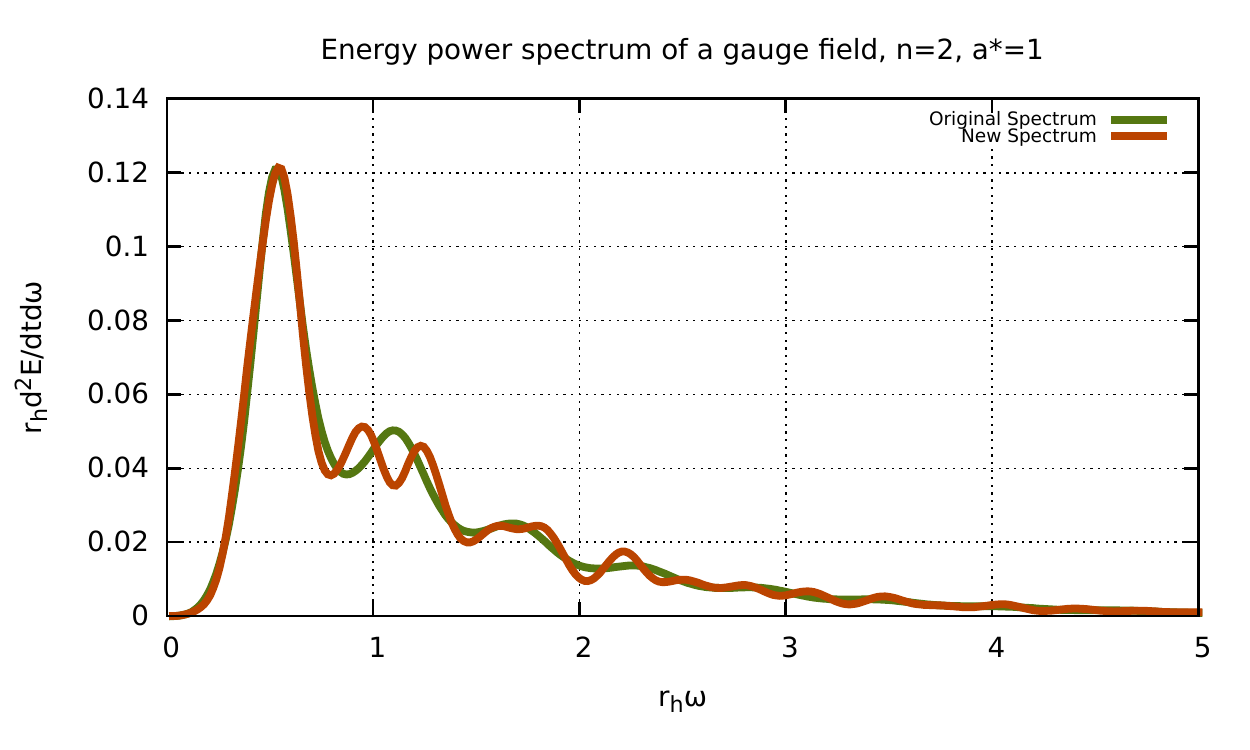}
\caption*{(a)}
\label{fig:2figsA}}
\qquad
\hspace{-1cm}\begin{minipage}{5cm}
\vspace{-0.25cm}
\includegraphics[width=5.8cm]{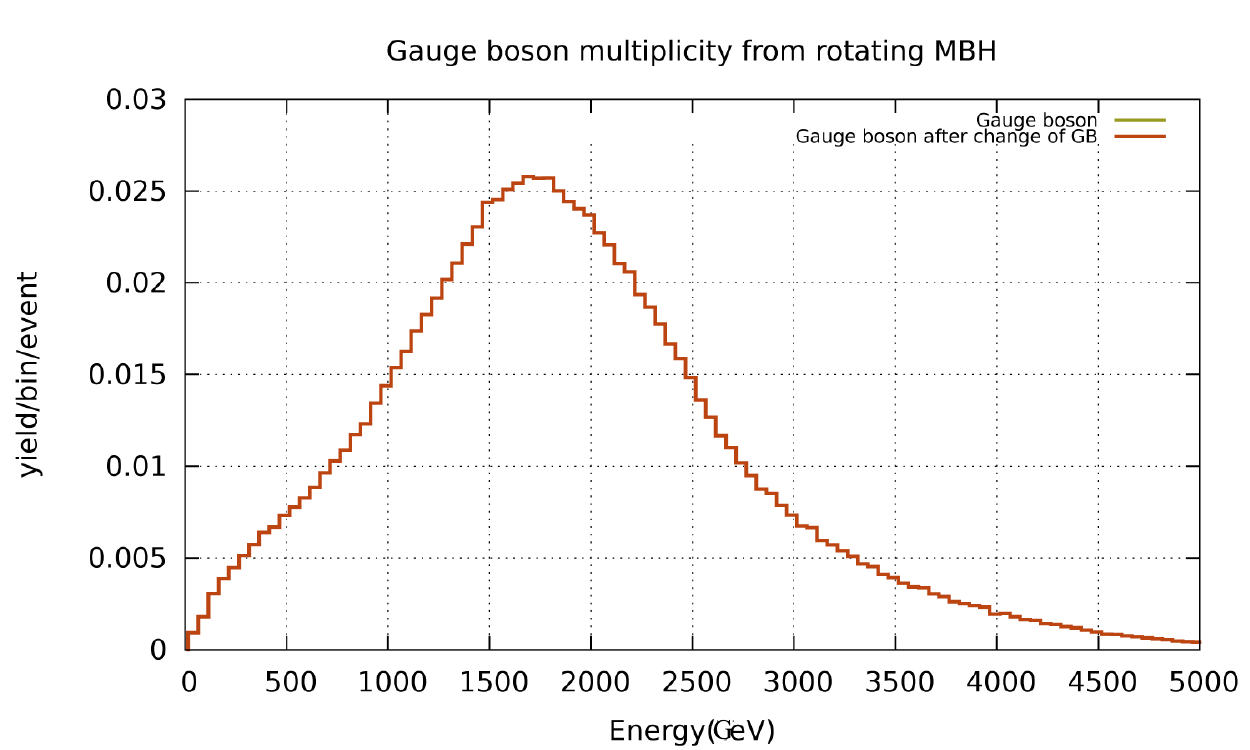}
\caption*{(b)}
\label{fig:2figsB}
\end{minipage}
\caption{(a) Energy power spectrum from a gauge field for $a_{\ast}=1$ before and after change of greybody factors. (b) gauge boson multiplicity before and after change of greybody factors. }
\label{f:group1}
\end{figure}
To test this sensitivity we changed by up to $15\%$, the grey body factors via the addition of a pseudo-random function in the BlackMax codes while we left the normalization unchanged. Then we investigated the final output of the BlackMax before and after this change of greybody factors for spinor and gauge  particles (because of the low statistic in the scalar case we were not interested to investigate them). Results of this study are shown in figures \ref{f:group1} and \ref{f:group}.
Figures \ref{f:group1}.a and \ref{f:group}.a represent the implemented change in the greybody factors for gauge bosons and spinor fields and figures \ref{f:group1}.b, \ref{f:group}.b and \ref{f:group}.c show the multiplicity of the gauge bosons, quarks and charged leptons before and after the change in greybody factors. As is clear, the multiplicity of the gauge bosons has not changed and the change of the quark and charged lepton multiplicities is negligible, indicating that BlackMax has low sensitivity to the exact form of the greybody factors. In this sense, in the absence of the grey body factors for graviton emission from a rotating MBH one may introduce an approximate profile for graviton emission and possibly study the final products of the MBH decay to see in particular if the graviton emission  increases with rotation and possibly competes with brane emission.
 
\section{Conclusion}
Micro black hole formation and evaporation is one of the possible consequences of low scale gravity models and the emitted Hawking radiation from an MBH will be dressed up by grey body factors. As a full numerical analysis of graviton emission from rotating black holes is not available, calculation of grey body factors for graviton emission is not possible. So, any approximate way to introduce graviton grey body factors can be valuable if we can be sure that incorporation of this approximate profile in the exiting MBH event generators might be a good alternative in the absence of exact profiles. For this purpose we studied the sensitivity of the BlackMax MBH event generators to the change of the grey body factors for different particle profiles and we found that this change has negligible affect on the final product of the MBH decay.

\end{document}